\documentclass[journal,transmag]{IEEEtran}
\usepackage{amsmath}
\usepackage{graphicx}
\usepackage{subcaption}
\usepackage{float}
\usepackage{multirow}
\usepackage[table,xcdraw]{xcolor}
\usepackage[noadjust]{cite}

\begin{document}
	\title{Time-Domain Analysis of PWM Inverters}
	\author{ \IEEEauthorblockN{Siddharth Tyagi and Isaak Mayergoyz, \IEEEmembership{Fellow,~IEEE}}
		\IEEEauthorblockA{Department of Electrical and Computer Engineering, University of Maryland, College Park, MD 20742 USA}
		\thanks{Corresponding author:  S. Tyagi, email: styagi@umd.edu}}
	\maketitle
	
	\begin{abstract}
 	The time-domain analysis of pulse width modulated (PWM) single-phase inverters is presented for different load circuits. It is demonstrated that this analysis can be reduced to the solution of linear simultaneous algebraic equations with two diagonal matrices. Analytical solutions of such equations are easily found which leads to the explicit expressions for the output voltages and currents in terms of switching time-instants. This technique is presented  for second and third order circuits, however it can be used in principle for any higher-order linear load circuit subject to pulse width modulated voltages. 
 	\end{abstract}
\newcommand{\vst}{v_s(t)}
\newcommand{\vct}{v_C(t)}
\newcommand{\ilt}{i_L(t)}
\newcommand{\ila}{i(t)}
\newcommand{\ilb}{i_1(t)}

\section*{\textbf{Introduction}}
The pulse width modulation (PWM) is ubiquitous in the design of various power electronics converters  \cite{holmes2003pulse, patrick2014fundamentals}. PWM is especially fundamental for the operation of inverters which are used for the frequency control of speed of ac motors \cite{holmes2003pulse, patrick2014fundamentals,buja1977optimal, denis2016core} (including spindle motors of hard disk drives \cite{jabbar2004analysis},\cite{jang2005bipolar}). PWM is also instrumental for the design of various uninterruptible power supplies (UPS) \cite{chen1995combination} as well as for the integration of renewal energy sources with the existing power grids \cite{carrasco2006power}. 

The principle of PWM is to generate voltages which are trains (sequences) of rectangular pulses. The widths of these pulses are properly modulated to suppress lower-order line voltage harmonics at the expense of higher-order harmonics, which are in turn suppressed by inductors in inverter circuits. Traditionally, the PWM is studied on the level of line-voltages while the effects of various loads on the time-variations of desired output voltages and currents are left unattended. 

In this paper, we are concerned with the analytical time-domain analysis of PWM for different load circuits used in applications. This time-domain analysis is useful for the assessment of  distortions in output voltages and currents caused by higher-order harmonics. The time-domain analysis is also needed for the calculation of hysteresis and eddy current losses in inductors \cite{mayergoyz1998nonlinear, mayergoyz1998nonlinear2, yao2017pwm, denis2016core} used for the suppression of high-order harmonics . This time-domain analysis is also fundamental for the development of optimal PWM schemes \cite{mayergoyz2018optimal, tyagi2020optimal} which result in minimization of the total harmonic distortion as well as possible selective harmonic elimination \cite{czarkowski2002solving}.  

In this paper, we perform the analytical time-domain analysis of PWM for single-phase inverters. The obtained results can be easily extended to the three-phase inverters. This can be done by using per-phase time-domain analysis \cite{tyagi2020optimal} of these inverters. The loads discussed in this paper are traditional $LR$ circuits as well as more sophisticated $L$-$RC$ and $L$-$C$-$LR$ circuits. Some of these circuits are used in UPS \cite{chen1995combination}. It is also demonstrated that these higher-order load circuits more efficiently suppress higher-order harmonics. Finally, the developed analytical technique can be in principle used for the analysis of any higher-order linear load circuit subject to PWM voltages.

\section*{\textbf{Technical Discussion}}
\begin{figure}[h!]
	\centering
	\includegraphics[trim={0 0.5cm 0 2cm}, width=1\linewidth]{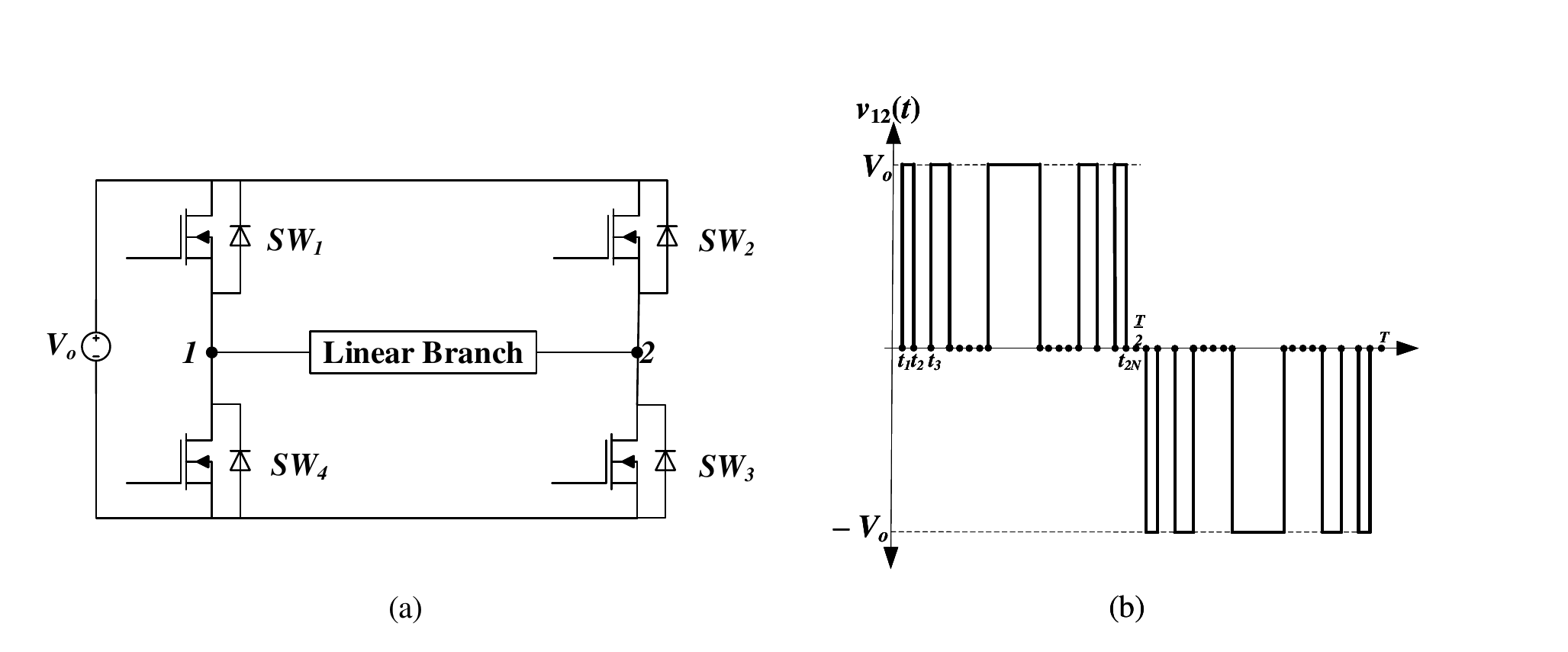}
	\caption{(a) Single-phase H-bridge inverter, (b) inverter voltage.}
	\label{fig1}
\end{figure}

Consider a single-phase inverter shown in Fig.~\ref{fig1}(a). Our goal is to derive the analytical expressions for output voltages or currents in the case when the loads are $L$-$RC$ and the $L$-$C$-$LR$ circuits shown in Fig.~\ref{LRC} and \ref{LCLR}, respectively, while the inverter voltage $v_{12}(t)$ is a train of rectangular pulses, as shown in Fig.~\ref{fig1}(b). Here, $T$ is the time-period , $\omega = \frac{2\pi}{T}$ is the frequency, and $N$ is the number of pulses in the half-period $0 < t < T/2$. This PWM voltage can be characterized by switching time-instants $t_1$, $t_2$, $...$, $t_{2N}$.

\section{Case of $L$-$RC$ load }
\begin{figure}[h!]
	\centering
	\includegraphics[trim={0 0.5cm 0 1.2cm}, width=0.8\linewidth]{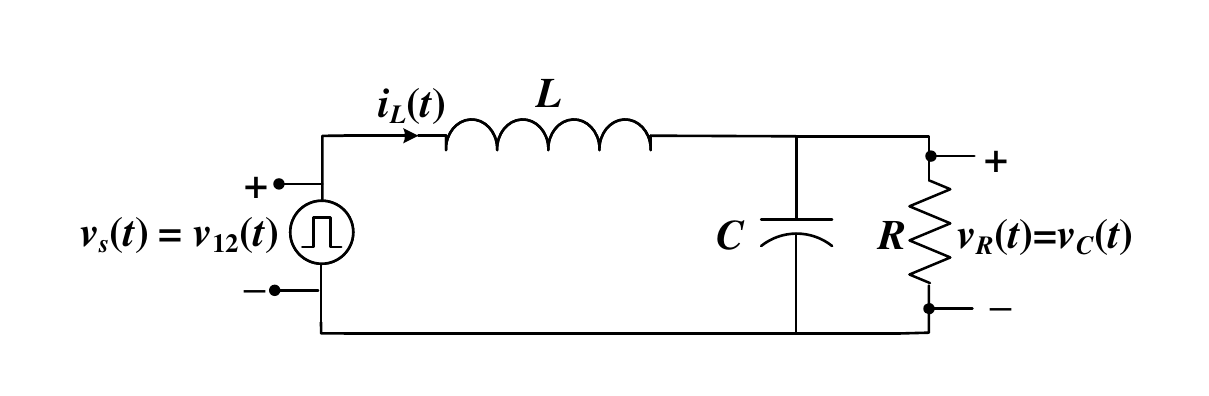}
	\caption{$L$-$RC$ circuit excited by PWM voltage.}
	\label{LRC}
\end{figure}

We intend to derive the analytical expression for the output voltage $v_R(t) = v_C(t)$ in Fig.~\ref{LRC}.

It is clear that the following formula is valid:
\begin{equation}
\vst=\begin{cases}
0, & \text{if $ t_{2j} < t < t_{2j+1} $},\\
V_o, & \text{if $ t_{2j+1} < t < t_{2j+2} $},
\end{cases} \label{vt}
\end{equation}
where $ j = 0,1,2, ... , N, $  and
\begin{equation} t_0 = 0, \; \; \;  t_{2N+1} = \frac{T}{2}. \label{tif}\end{equation}
It is apparent that the voltage $v_s(t)$ has the half-wave symmetry:
\begin{equation}
v_s \left( t + \frac{T}{2}\right) = -v_s(t).\label{hws}  
\end{equation}

By using Kirchhoff voltage law (KVL)  and Kirchhoff current law (KCL) the following differential equations can be written: 
\begin{align}
L \frac{d\ilt}{dt} + \vct &= \vst, \label{KVL}\\
C \frac{d\vct}{dt} + \frac{\vct}{R} &= \ilt. \label{KCL}
\end{align}
Differentiating equation (\ref{KCL}) and then substituting it in (\ref{KVL}), we obtain:
\begin{equation}
LC \frac{d^2\vct}{dt^2} + \frac{L}{R} \frac{d\vct}{dt}+ \vct = \vst. \label{vceq}
\end{equation}
The general solution for the corresponding homogeneous equation can be written in the form:
\begin{equation}
v_{C}^h(t) = Ae^{s_1t} + Be^{s_2t}, \label{homoLRC}
\end{equation} 
where $s_1$ and $s_2$ are roots of the following quadratic equation:
\begin{equation}
LC s^2 + \frac{L}{R}s + 1 = 0. \label{diffLRC}
\end{equation}
Here, for the sake of simplicity we consider the generic case when the roots of the above equation are distinct.

The particular solution to equation (\ref{vceq}), when $v_s(t) = V_o$ is given by the formula:
\begin{equation}
v_{C}^p(t) = V_o. \label{part}
\end{equation}
Thus, by using equations (\ref{homoLRC}) and (\ref{part}), we arrive at the following formula:
\begin{equation}
\resizebox{0.485\textwidth}{!}{$ v_C(t)= \begin{cases}
	A_{2j+1}e^{s_1t} + B_{2j+1}e^{s_2t}, & \text{if $ t_{2j} < t < t_{2j+1} $},\\
	A_{2j+2}e^{s_1t} + B_{2j+2}e^{s_2t} + V_o, & \text{if $ t_{2j+1} < t < t_{2j+2} $},
	\end{cases} $}\label{vcsol1}
\end{equation}
where  $ j = 0,1,2, ... , N$. The last formula can be rewritten as:
\begin{equation}
\vct = A_{k}e^{s_1t} + B_{k}e^{s_2t} + \chi_{k},  \text{if $ t_{k-1} < t < t_{k} $},  \label{vcsol} 
\end{equation}
where 
\begin{equation}
	\chi_k = \frac{V_o}{2} \cdot \left( 1 + (-1)^k\right), \text{ for }  k = 1, 2, ... , 2N+1.  \label{chi}
\end{equation}

It is clear that $\chi_k$ takes two values: $0$ or $V_o$. Furthermore, from formula (11) we find:
\begin{equation}
\frac{dv_C(t)}{dt} = s_1 A_{k}e^{s_1t} + s_2 B_{k}e^{s_2t},  \text{ if }  t_{k-1} < t < t_{k}. \label{vcderv}
\end{equation}
We now proceed to find the expressions for the coefficients $A_k$ and $B_k$. This is done on the basis of the following conditions: 
\begin{enumerate}
	\item continuity of voltage $\vct$:
	\begin{align}
	v_C(t_{k}^-) &= v_C(t_{k}^+), \label{con1}
	\end{align}
	\item continuity of the derivative of the voltage $\vct$:
	\begin{align}
	\frac{dv_C}{dt}(t_{k}^-) &= \frac{dv_C}{dt}(t_{k}^+), \label{con2}
	\end{align}
	\item half-wave anti-periodic boundary conditions:
	\begin{equation}
	v_{C} (0) = - v_{C} \left( \frac{T}{2}\right), \label{con3}
	\end{equation}
	\begin{equation}
	\frac{dv_{C}}{dt} (0) = - \frac{dv_{C}}{dt}\left( \frac{T}{2}\right). \label{con4}
	\end{equation}
\end{enumerate}
The validity of formula (\ref{con2}) follows from equation (\ref{KCL}) and continuity of $\vct$ and $\ilt$. Similarly, formula (\ref{con4}) follow from (\ref{KCL}), (\ref{con3})  and the half-wave periodicity of $\ilt$. 

By using equations (\ref{vcsol}), (\ref{vcderv}), (\ref{con1}) and (\ref{con2}), we can write
\begin{align}
A_ke^{s_1t_k} + B_ke^{s_2t_{k}} + \chi_k &= A_{k+1}e^{s_1t_{k}} + B_{k+1}e^{s_2t_{k}} + \chi_{k+1}, \label{e1} \\
s_1A_ke^{s_1t_k} + s_2B_ke^{s_2t_{k}} &= s_1A_{k+1}e^{s_1t_{k}} + s_2B_{k+1}e^{s_2t_{k}}, \label{e2}
\end{align}
where $ k = 1, 2, ... , 2N $. 

Formulas (\ref{e1}) and (\ref{e2}) define a set of simultaneous linear equations with a four-diagonal matrix. It is remarkable that explicit analytical solutions to these equations can be found. This is because these equations have a special mathematical structure which allows to reduce them to two sets of decoupled simultaneous equations with two-diagonal matrices for coefficients $A_k$ and $B_k$, respectively. Indeed, by multiplying equation (\ref{e1}) with $s_2$ and subtracting from (\ref{e2}) we find:
\begin{equation}
(s_1 - s_2) A_ke^{s_1t_k} - s_2\chi_k = (s_1-s_2)A_{k+1}e^{s_1t_k} - s_2\chi_{k+1},
\end{equation}
which can then be rearranged as follows:
\begin{equation}
A_{k} - A_{k+1} = \frac{s_2}{s_1 - s_2} \left[ \chi_{k} - \chi_{k+1}\right]e^{-s_1t_k}.
\end{equation}
It is apparent from equation (\ref{chi}) that
\begin{equation}
\chi_{k} - \chi_{k+1} = (-1)^k V_o.  
\end{equation}
Thus, from the last two equations, we obtain: 
\begin{equation}
A_{k} - A_{k+1} = (-1)^k V_o \frac{s_2}{s_1 - s_2}  e^{-s_1t_k}. \label{Akall}
\end{equation}
Thus, it is evident that equations (\ref{Akall}) are simultaneous linear equations with two-diagonal matrix. Now, if we add equations (\ref{Akall}) for $k =1, 2,.. , 2N$, we obtain
\begin{equation}
A_1 - A_{2N+1} = V_o\frac{s_2}{s_1 - s_2} \sum_{n = 1}^{2N} (-1)^n e^{-s_1t_n}.  \label{Afl1}
\end{equation}
Next, from equations (\ref{vcsol}), (\ref{vcderv}) and the boundary conditions (\ref{con3})-(\ref{con4}), we find:
\begin{align}
A_1 + B_1  &= - A_{2N+1}e^{\frac{T}{2}s_1} - B_{2N+1}e^{\frac{T}{2}s_1}, \label{e3} \\
s_1A_1 + s_2B_1 &= - s_1A_{2N+1}e^{\frac{T}{2}s_1} - s_2B_{2N+1}e^{\frac{T}{2}s_1}, \label{e4}
\end{align}

Again, by multiplying equation (\ref{e3}) by $s_2$ and then subtracting it from (\ref{e4}), we obtain:
\begin{equation}
A_1 + A_{2N+1}e^{\frac{T}{2}s_1} = 0.  \label{Afl2}
\end{equation}
By solving simultaneous equations (\ref{Afl1}) and (\ref{Afl2}) we find:
\begin{align}
A_1 &=    \frac{ V_o s_2 e^{ \frac{T}{2}s_1}}{(s_1 - s_2) \left( 1+e^{\frac{T}{2}s_1}\right)}  \sum_{n = 1}^{2N} (-1)^n e^{-s_1t_n}, \label{final1} \\
A_{2N+1} &= - \frac{ V_o s_2}{(s_1 - s_2) \left( 1+e^{\frac{T}{2}s_1}\right)}  \sum_{n = 1}^{2N} (-1)^n e^{-s_1t_n}. \label{final2}
\end{align}
Having found $A_1$, the other coefficients $A_k$ can be computed as follows:
\begin{equation}
A_k = A_1 -  \frac{s_2 V_o }{s_1 - s_2} \sum_{n = 1}^{k-1} (-1)^n e^{-s_1t_n}. \label{final3}
\end{equation}
The last expression is derived by adding the first $(k-1)$ equations defined by formula (\ref{Akall}). 
Finally, by using permutational symmetry of equations (\ref{e1}) and (\ref{e2}) similar expressions can be immediately found for the coefficients $B_k$, by interchanging $s_1$ and $s_2$ in (\ref{final1})-(\ref{final3}). Thus, by using these analytical expressions for $A_k$ and $B_k$ in equation (\ref{vcsol1}), we arrive at the analytical formula for $v_C(t)$ which is valid for any choice of switching time-instants $t_k$.

\section{Case of $L$-$C$-$LR$ load}
\begin{figure}[h!]
	\centering
	\includegraphics[trim={0 0.5cm 0 1.2cm}, width=0.8\linewidth]{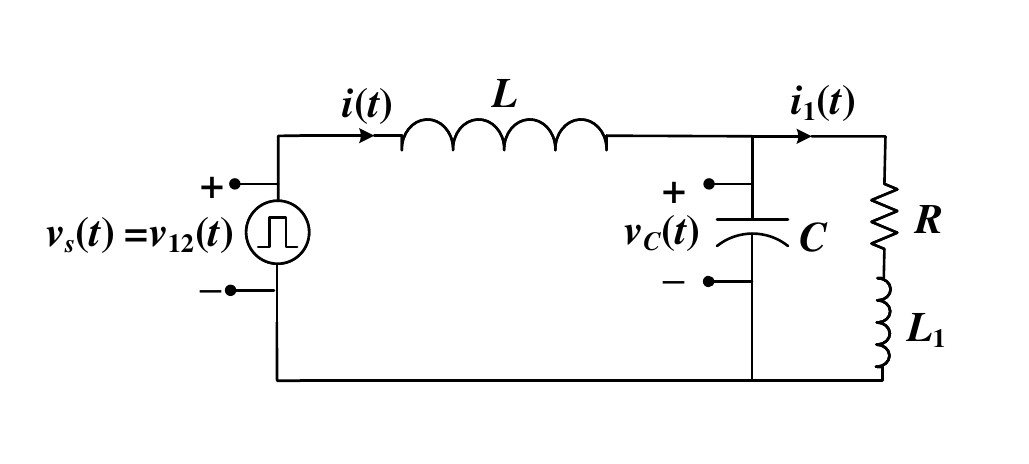}
	\caption{$L$-$C$-$RL$ circuit excited by PWM voltage.}
	\label{LCLR}
\end{figure}

In this section, we extend the previous analysis to the case of the circuit shown in Fig.~\ref{LCLR}. We intend to derive the analytical expression for the output current $\ilb$ in Fig.~\ref{LCLR}. As before, the voltage $v_s(t)$ is described by equation (\ref{vt}) and it satisfies the half-wave symmetry condition (\ref{hws}).

By using KVL and KCL the following differential equations can be written:  
\begin{align}
\frac{di(t)}{dt}&= -\frac{1}{L} \vct + \frac{1}{L} \vst, \label{KVL1}\\ 
\frac{di_{1}(t)}{dt}&= -\frac{R}{L_1} \ilb + \frac{1}{L_1} \vct, \label{KVL2} \\
\frac{dv_C(t)}{dt} &= \frac{1}{C}\ila - \frac{1}{C} \ilb. \label{KCL2}
\end{align}
Next, we define the state-vector
\begin{equation}
\mathbf{x}(t) = \begin{bmatrix}
\ila \\ \ilb \\ \vct 
\end{bmatrix}, \label{defx}
\end{equation}
and represent the equations (\ref{KVL1})-(\ref{KCL2}) in state-variable form:
\begin{equation}
\dot{\mathbf{x}}(t)= \begin{bmatrix}
0 & 0 & -\frac{1}{L} \\
0 & -\frac{R}{L_1} & \frac{1}{L_1} \\
\frac{1}{C} & -\frac{1}{C} & 0
\end{bmatrix} \mathbf{x}(t) + \begin{bmatrix}
\frac{1}{L} \\ 0 \\ 0 
\end{bmatrix} \vst. \label{stvar}
\end{equation}
The characteristic equation for the matrix in formula (\ref{stvar}) can be written as follows: 
\begin{equation}
s^3 + \frac{R}{L_1}s^2 + \frac{1}{C}\left( \frac{1}{L}+ \frac{1}{L_1}\right)s + \frac{R}{LL_1C}=0.
\end{equation}
For the sake of simplicity, we consider the generic case when the three roots $s_1, s_2,$ and  $s_3$ of the above equation are distinct. This implies that for $\ilb$, the general solution of the corresponding homogeneous equation has the form:
\begin{equation}
i_1^h(t) = Ae^{s_1t} + Be^{s_2t} + Ce^{s_3t}. \label{homob}
\end{equation}
On the other hand, it is easy to see that the particular solution to the state-equation (\ref{stvar}), for the time-intervals when $v_s(t)=V_o$ can be written as follows:
\begin{equation}
i^p(t) = i_1^p(t) = \frac{V_o}{R}, \hspace{20pt} v_{C}^p(t) = V_o.  \label{partb}
\end{equation}
Using (\ref{homob}) and (\ref{partb}), the general solution for $\ilb$ can be expressed as:
 \begin{equation}
\resizebox{0.49\textwidth}{!}{$i_1(t)= \begin{cases}
	A_{2j+1}e^{s_1t} + B_{2j+1}e^{s_2t}+C_{2j+1}e^{s_3t}, & \text{if $ t_{2j} < t < t_{2j+1} $},\\
	A_{2j+2}e^{s_1t} + B_{2j+2}e^{s_2t} +C_{2j+2}e^{s_3t}+ \frac{V_o}{R}, & \text{if $ t_{2j+1} < t < t_{2j+2} $}.
	\end{cases} $}\label{ilbsol1}
\end{equation}
where  $ j = 0,1,2, ... , N$. The above equation can be represented in the form:
\begin{equation}
\ilb = A_{k}e^{s_1t} + B_{k}e^{s_2t}+ C_{k}e^{s_3t} + \tilde{\chi}_{k},  \text{if $ t_{k-1} < t < t_{k} $},  \label{ilbsol} \\
\end{equation}
where
\begin{equation}
\tilde{\chi}_k = \frac{V_o}{2R} \cdot \left( 1 + (-1)^k\right), \label{tchi}
\end{equation}
for $ k = 1, 2, ... , 2N+1 $. It is apparent that $\tilde{\chi}_{k}$ takes two values: $0$ or $\frac{V_o}{R}$. 

It is clear that the following formulas can be written for the derivatives of $\ilb$:
\begin{equation}
\frac{di_1(t)}{dt} = s_1 A_{k}e^{s_1t} + s_2 B_{k}e^{s_2t} +s_3 C_{k}e^{s_3t} ,  \text{if $ t_{k-1} < t < t_{k} $},\label{ilbfderv}
\end{equation}
\begin{equation}
\frac{d^2i_1(t)}{dt^2} = s_1^2 A_{k}e^{s_1t} + s_2^2 B_{k}e^{s_2t} +s_3^2 C_{k}e^{s_3t} ,  \text{if $ t_{k-1} < t < t_{k} $}. \label{ilbsderv}
\end{equation}

We now proceed to derive the expressions for the unknown coefficients $A_k$, $B_k$ and $C_k$. This is done by using the continuity condition for $\ilb$ as well as for its first and second-order derivatives. It is clear that the state-variables $\ila, \ilb$ and $\vct$ are continuous. Moreover, it follows from equation (\ref{KVL2}) and (\ref{KCL2}), the derivatives of $\ilb$ and $\vct$ are also continuous. Furthermore, by differentiating equation (\ref{KVL2}), we can also conclude that the second-order derivative of $\ilb$ is continuous. Thus, by applying the above stated continuity conditions at $t_k$ for all $ k = 1, 2, ... , 2N$, the following equations can be derived using (\ref{ilbsol}), (\ref{ilbfderv}) and (\ref{ilbsderv}):
\begin{align}
&A_{k}e^{s_1t_k} + B_{k}e^{s_2t_k}+ C_{k}e^{s_3t_k} + \tilde{\chi}_{k} \nonumber \\=  &A_{k+1}e^{s_1t_k} + B_{k+1}e^{s_2t_k}+ C_{k+1}e^{s_3t_k} + \tilde{\chi}_{k+1}, \label{eq1} 
\end{align}
\begin{align}
&s_1 A_{k}e^{s_1t_k}+ s_2 B_{k}e^{s_2t_k} +s_3 C_{k}e^{s_3t_k} \nonumber \\ &= 
s_1 A_{k+1}e^{s_1t_k}+ s_2 B_{k+1}e^{s_2t_k} +s_3 C_{k+1}e^{s_3t_k}, \label{eq2}
\end{align}
\begin{align}
&s_1^2 A_{k}e^{s_1t_k}+ s_2^2 B_{k}e^{s_2t_k} +s_3^2 C_{k}e^{s_3t_k} \nonumber \\ &= 
s_1^2 A_{k+1}e^{s_1t_k}+ s_2^2 B_{k+1}e^{s_2t_k} +s_3^2 C_{k+1}e^{s_3t_k}.\label{eq3}
\end{align}
These are linear coupled equations for the coefficients $A_k$, $B_k$ and $C_k$ with the six diagonal matrix. It is remarkable that these equations can be decoupled and reduced to three separate sets of simultaneous linear equations with two-diagonal matrices. This is accomplished as follows. Multiplying (\ref{eq1}) by $s_2$ and subtracting it from (\ref{eq2}), we obtain:
\begin{align}
&(s_1 - s_2) A_{k}e^{s_1t_k} + (s_3 - s_2) C_{k}e^{s_3t_k}-s_2\tilde{\chi}_{k}= \nonumber \\
&(s_1 - s_2) A_{k+1}e^{s_1t_k} + (s_3 - s_2) C_{k+1}e^{s_3t_k}-s_2\tilde{\chi}_{k+1} \label{eq4}
\end{align} 

Similarly, multiplying (\ref{eq1}) by $s_2^2$ and subtracting it from (\ref{eq3}), we get:
\begin{align}
&(s_1^2 - s_2^2) A_{k}e^{s_1t_k} + (s_3^2 - s_2^2) C_{k}e^{s_3t_k}-s_2^2\tilde{\chi}_{k} \nonumber \\ = 
&(s_1^2 - s_2^2) A_{k+1}e^{s_1t_k} + (s_3^2 - s_2^2) C_{k+1}e^{s_3t_k}-s_2^2\tilde{\chi}_{k+1} \label{eq5}
\end{align}
It is clear that coefficients $B_k$  have been eliminated. We next eliminate coefficients $C_k$ by multiplying (\ref{eq4}) by $(s_2+s_3)$ and subtracting it from (\ref{eq5}). After simple algebraic transformations, the following equations for $A_k$ emerge: 
\begin{equation}
A_k - A_{k+1} = \frac{s_2s_3}{(s_1-s_2)(s_1-s_3)} \left[ \tilde{\chi}_{k+1}- \tilde{\chi}_{k} \right] e^{-s_1t_k}
\end{equation}
for $k = 1, 2, ... 2N$.
 It is clear from (\ref{tchi}) that:
\begin{equation}
\tilde{\chi}_{k+1}- \tilde{\chi}_{k} = (-1)^{k+1} \frac{V_o}{R}.
\end{equation} 
Thus, from the last two formulas, we conclude :
\begin{equation}
A_k - A_{k+1} =\frac{V_o}{R} \frac{s_2s_3}{(s_1-s_2)(s_1-s_3)} (-1)^{k+1} e^{-s_1t_k}. \label{eqnn}
\end{equation}
By adding the last equation written for $k = 1, 2, ... , 2N$, we arrive at the following formula:
\begin{equation}
A_1 - A_{2N+1} = \frac{V_o}{R}\frac{s_2s_3}{(s_1-s_2)(s_1-s_3)} \sum_{n = 1}^{2N}(-1)^{n+1} e^{-s_1t_n}. \label{aif1}
\end{equation}

Next, the following half-wave anti-periodic boundary conditions for $\ilb$ as well as its first and second order derivatives can be established:
\begin{align}
i_1(0) &= - i_1\left( \frac{T}{2}\right), \label{hw1}\\
\frac{di_1}{dt}(0) &= - \frac{di_1}{dt}\left( \frac{T}{2}\right), \label{hw2}\\
\frac{d^2i_1}{dt^2}(0) &= - \frac{d^2i_1}{dt}\left( \frac{T}{2}\right). \label{hw3}
\end{align}
The above equations follow from formula (\ref{hws}) which implies the half-wave symmetry of the state variables and their derivatives. 
Using the above equations along with (\ref{tif}), (\ref{ilbsol}), (\ref{ilbfderv}) and (\ref{ilbsderv}) we obtain:
\begin{align}
&A_{1}+ B_{k} + C_{k}  \nonumber \\ &= - A_{2N+1}e^{s_1\frac{T}{2}} - B_{2N+1}e^{s_2\frac{T}{2}} - C_{2N+1}e^{s_3\frac{T}{2}}, \label{eq6} 
\end{align}
\begin{align}
&s_1A_{1}+ s_2B_{k} +s_3 C_{k}  \nonumber \\ &= - s_1A_{2N+1}e^{s_1\frac{T}{2}} - s_2B_{2N+1}e^{s_2\frac{T}{2}} - s_3C_{2N+1}e^{s_3\frac{T}{2}}, \label{eq7}
\end{align}
\begin{align}
&s_1^2A_{1}+ s_2^2B_{k} +s_3^2 C_{k}  \nonumber \\ &= - s_1^2A_{2N+1}e^{s_1\frac{T}{2}} - s_2^2B_{2N+1}e^{s_2\frac{T}{2}} - s_3^2C_{2N+1}e^{s_3\frac{T}{2}} .\label{eq8}
\end{align}
Again, using the same steps as before, we can derive the following equation:
\begin{equation}
A_1 + A_{2N+1}e^{s_1\frac{T}{2}}=0. \label{eq9}
\end{equation}
Finally, we can solve the simultaneous equations (\ref{aif1}) and (\ref{eq9}). This yields the following result:
\begin{align}
A_1 &=  \frac{V_o s_2s_3}{R (s_1-s_2)(s_1-s_3)\left( 1+e^{-s_1\frac{T}{2}} \right)} \sum_{n = 1}^{2N}(-1)^{n+1} e^{-s_1t_n}. \label{final1LCLR}
\end{align} 
Having obtained $A_1$, all other coefficients $A_k$ can be computed according to the formula: 
\begin{equation}
A_k = A_1 - \frac{V_os_2s_3}{R(s_1-s_2)(s_1-s_3)} \sum_{n = 1}^{k-1}(-1)^{n+1} e^{-s_1t_n}. \label{final2LCLR}
\end{equation}
This formula is obtained by adding the first $(k-1)$ equations of the form (\ref{eqnn}). 
By using the symmetry argument, similar expressions for coefficients $B_k$ and $C_k$ can be immediately written by the appropriate interchanging (permutation) of $s_1$, $s_2$ and $s_3$. In this way, the explicit analytical expression for the current $i_1(t)$ is obtained. 

\section{Numerical Results}
Here, we present computational results for the output currents through the resistors for the $L$-$RC$ and $L$-$C$-$LR$ loads using the expressions derived in the previous sections. We compare these results with the output currents in the case of the $LR$ loads for which the analytical expressions have been previously reported \cite{patrick2014fundamentals, tyagi2020optimal}. The values of the parameters are chosen as follows: $V_o = 100$ V, $f = 60$ Hz $N=11$. The switching time-instants $t_1$, $t_2$, $...$. $t_{2N}$ can be computed using traditional sinusoidal PWM, where pulse widths are modulated proportional to the desired sinusoidal value, and placed in the center of each switching time interval. 

\begin{figure}[h!]
 	\centering
 	\includegraphics[trim={0 0.5cm 0 1.2cm}, width=0.89\linewidth]{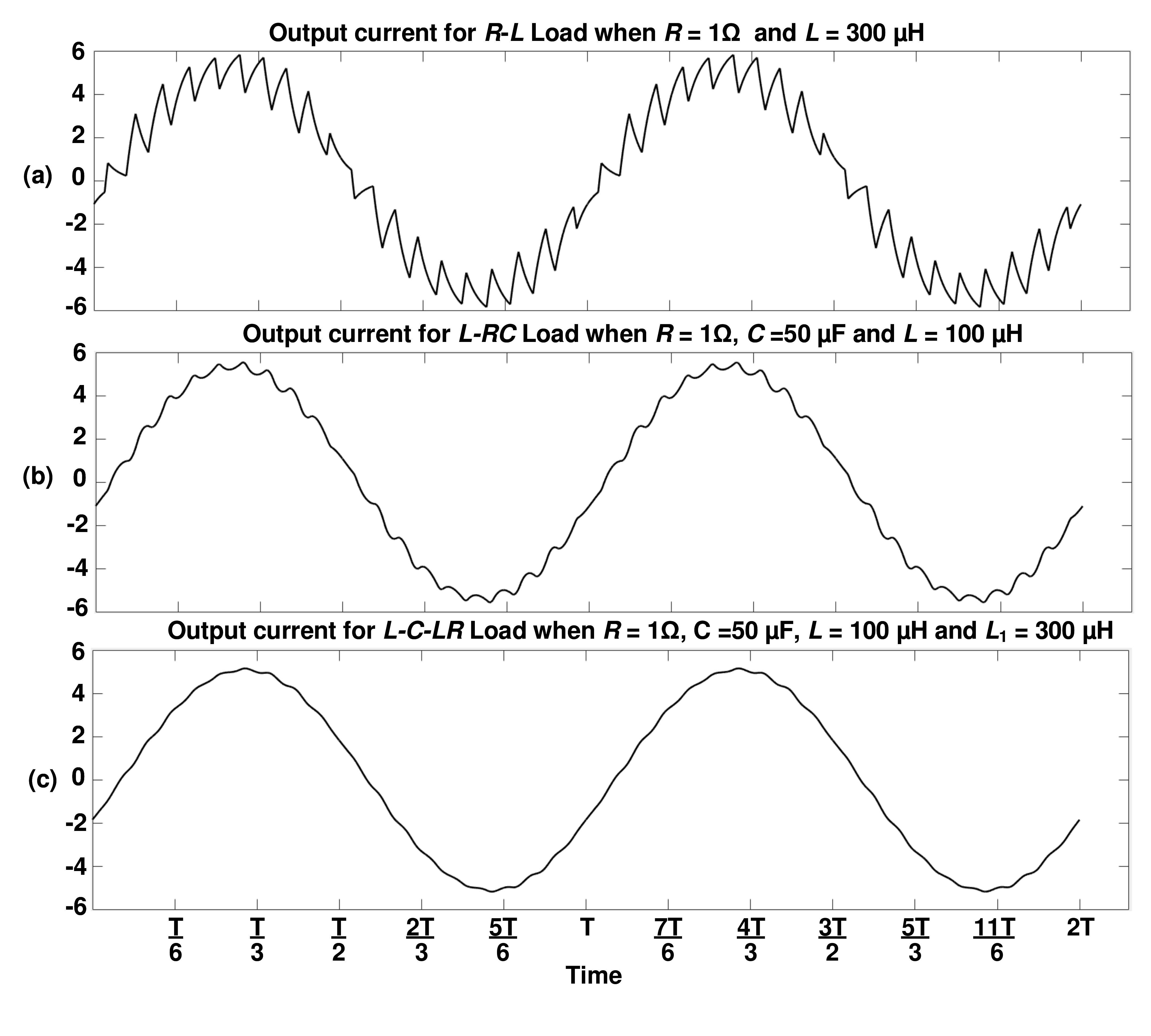}
 	\caption{Comparison of output currents for the $LR$, $L$-$RC$ and $L$-$C$-$LR$ loads.}
 	\label{NR1}
 \end{figure}

Fig.~\ref{NR1} presents the output currents through the resistor computed for the $LR$, $L$-$RC$ and $L$-$C$-$LR$ load circuits.  For each case, the load resistor $R = 1 \Omega$. For the $L$-$RC$ and $L$-$C$-$LR$ loads, the capacitor $C = 50$ $\mu$F and $L$ = $100$  $\mu$H. The value of the output inductance for the $LR$ and the $L$-$C$-$LR$ loads was chosen to be $300$ $\mu$H. The presented results clearly reveal the significant reduction in the output current waveform ripple as the order of the load circuits increases. 

\begin{figure}[h!]
	\centering
	\includegraphics[trim={0 0.5cm 0 1.2cm}, width=0.85\linewidth]{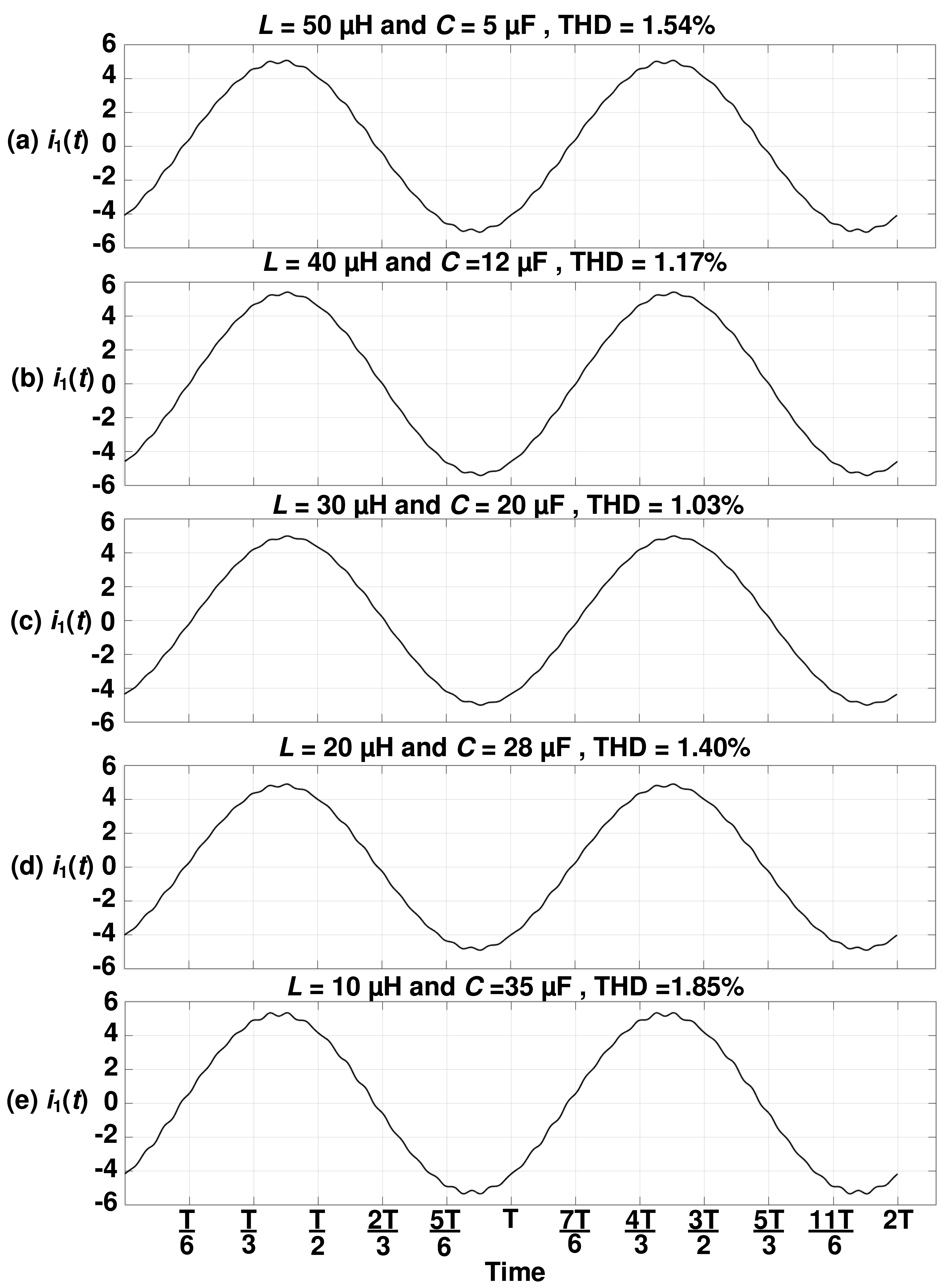}
	\caption{Comparison of output currents for $i_1(t)$ for the $L$-$C$-$LR$ loads, for different values of parameters $L$ and $C$. These results illustrate that better (or comparable) harmonic content in output currents is possible using a lower value of inductance.}
	\label{NR2}
\end{figure}

In Table \ref{Tab1} we present numerical results for the $L$-$C$-$LR$ load which reveal that lesser (or comparable) harmonic distortion in the output current can be achieved even with a lower value of inductance $L$, by selecting appropriate values of the capacitance $C$. The corresponding output current waveforms are shown in Fig.~\ref{NR2}. Here, the value of $L$ is reduced from $50$ $\mu$H to $10$ $\mu$H, and the total harmonic distortion (THD) \cite{holmes2003pulse, tyagi2020optimal} of the output current is computed. We observe that even when the value of $L$ is lowered from $50$ $\mu$H to $20$ $\mu$H the THD can be reduced by appropriate choice of capacitance $C$. Since inductors are bulky components, and also have associated hysteresis and eddy current losses, a reduction in inductance can help make inverter circuits lighter and more efficient. 
\begin{table}[h!]
	\centering
		\caption{Comparison of THD in output current of the $L$-$C$-$LR$ circuit, for fixed load of $R$=$1\Omega$ and $L_1$ = $300$ $\mu$H and different values of the parameters $L$ and $C$ (see Fig.~\ref{LCLR}).}
	\begin{tabular}{|c|c|c|}
		\hline
		\multicolumn{2}{|c|}{\begin{tabular}[c]{@{}c@{}}Values of \\ Parameters\end{tabular}} & \multirow{2}{*}{\begin{tabular}[c]{@{}c@{}}THD in output current\\   (in \%)\end{tabular}} \\ \cline{1-2}
		$L$ (in $\mu$H)                                 & $C$ (in $\mu$F)                                 &                                                                          \\ \hline
		50                                        & 5                                         & 1.54                                                                  \\
		40                                        & 12                                         & 1.17                                                                    \\
		30                                        & 20                                        & 1.03                                                                   \\
		20                                         & 28                                       & 1.40                                                                    \\
		10                                         & 35                                        & 1.83                                                                   \\ \hline
	\end{tabular}
\label{Tab1}

\end{table}

\bibliographystyle{ieeetran}
\bibliography{Referen2}

\end{document}